\documentstyle[twocolumn,prl,aps]{revtex}

\begin{document}
\draft
\title{Inclusive Electron Scattering from Nuclei at $x \simeq 1$}

\author{
J.~Arrington,$^1$\,
P.~Anthony,$^2$\,
R.~G.~Arnold,$^3$\,
E.~J.~Beise,$^1${$^,$}{\cite{atmaryland}}\,
J.~E.~Belz,$^1${$^,$}{\cite{atboulder}}\,
P.~E.~Bosted,$^3$\,
H.-J.~Bulten,$^4$\,
M.~S.~Chapman,$^5$\,
K.~P.~Coulter,$^6${$^,$}{\cite{atmich}}\,
F.~Dietrich,$^2$\,
R.~Ent,$^5${$^,$}{\cite{atcebaf}}\,
M.~Epstein,$^7$\,
B.~W.~Filippone,$^1$\,
H.~Gao,$^1${$^,$}{\cite{atillinois}}\,
R.~A.~Gearhart,$^8$\,
D.~F.~Geesaman,$^6$\,
J.-O.{} Hansen,$^5$\,
R.~J.~Holt,$^6$\,
H.~E.~Jackson,$^6$\,
C.~E.~Jones,$^4${$^,$}{\cite{atanl}}\,
C.~E.~Keppel,$^3${$^,$}{\cite{atvirgin}}\,
E.~R.~Kinney,$^{9}$\,
S.~Kuhn,$^{10}${$^,$}{\cite{atodu}}\,
K.~Lee,$^5$\,
W.~Lorenzon,$^1${$^,$}{\cite{atpenn}}\,
A.~Lung,$^3${$^,$}{\cite{atcaltech}}\,
N.~C.~R.~Makins,$^5$\,
D.~J.~Margaziotis,$^7$\,
R.~D.~McKeown,$^1$\,
R.~G.~Milner,$^5$\,
B.~Mueller,$^1$\,
J.~Napolitano,$^{11}$\,
J.~Nelson,$^5${$^,$}{\cite{atslac}}\,
T.~G.~O'Neill,$^1${$^,$}{\cite{atanl}}\,
V.~Papavassiliou,$^6${$^,$}{\cite{atillinoistech}}\,
G.~G.~Petratos,$^8${$^,$}{\cite{atkent}}\,
D.~H.~Potterveld,$^6$\,
S.~E.~Rock,$^3$\,
M.~Spengos,$^3$\,
Z.~M.~Szalata,$^3$\,
L.~H.~Tao,$^3$\,
K.~van~Bibber,$^2$\,
J.~F.~J.~van~den~Brand,$^4$\,
J.~L.~White,$^3$\,
D.~Winter,$^5${$^,$}{\cite{atcolombia}}\,
B.~Zeidman$^6$}

\address{\hss\\
$^1$\ California Institute of Technology, Pasadena, California
91125\\
$^2$\ Lawrence Livermore National Laboratory, Livermore, California
94550\\
$^3$\ American University, Washington, D. C. 20016\\
$^4$\ University of Wisconsin, Madison, Wisconsin 53706\\
$^5$\ Massachusetts Institute of Technology, Cambridge,
Massachusetts 02139\\
$^6$\ Argonne National Laboratory, Argonne, Illinois 60439\\
$^7$\ California State University, Los Angeles, California 90032\\
$^8$\ Stanford Linear Accelerator Center, Stanford, California
94309\\
$^{9}$\ University of Colorado, Boulder, Colorado 80309\\
$^{10}$\ Stanford University, Stanford, California 94305\\
$^{11}$\ Rensselaer Polytechnic Institute, Troy, New York 12180\\
\hss\\}

\date{\today}
\maketitle
\bigskip
\bigskip
\bigskip

\begin{abstract}
The inclusive $A(e,e')$ cross section for $x \simeq 1$ was measured on $^2$H,
C, Fe, and Au for momentum transfers $Q^2$ from  1$-$7 (GeV/c)$^2$.  The
scaling
behavior of the data was examined in the region of transition from $y$-scaling
to $x$-scaling.  Throughout this transitional region, the data exhibit
$\xi$-scaling, reminiscent of the Bloom-Gilman duality seen in free nucleon
scattering.

\end{abstract}

\narrowtext

In inclusive electron scattering, scaling functions are
important in the study of constituent sub-structure and interactions.  Scaling
is typically a sign that a simple reaction mechanism dominates the process,
allowing one to extract information on structure in a model independent way.
%Scaling has been observed for both quasielastic ($y$-scaling) and deep
%inelastic ($x$-scaling) electron scattering.  However, both of these have
%limited ranges of application.  This paper will examine transition from
%$y$-scaling to $x$-scaling, as well as scaling with respect to a `composite'
%scaling variable, Nachtmann $\xi$.
The most familiar scaling occurs in the limit of large energy transfer $\nu$
and momentum transfer $Q^2$ where the deep inelastic structure functions
$MW_1(\nu ,Q^2)$ and $\nu W_2(\nu ,Q^2)$ become functions only of the Bjorken
$x=Q^2/2M\nu$, where $M$ is the nucleon mass.  In this limit, $x$ can be
interpreted
as the longitudinal momentum fraction of the struck quark, and $MW_1$ and $\nu
W_2$ are related to the quark longitudinal momentum distribution.  Violations
of Bjorken scaling in the free nucleon exist at low $Q^2$ due to target mass
and higher-twist effects.  To correct for the effects of target mass at finite
$Q^2$, the Nachtmann variable $\xi = 2x/[1+(1+4M^2x^2/Q^2)^{1/2}]$ has been
used in place of $x$.  This has been shown to be the correct variable in which
to study the logarithmic QCD scaling violations in the nucleon\cite{hg76}.

Similarly, for quasielastic electron-nucleus scattering at high momentum
transfer, the `reduced' cross section was predicted\cite{west75}, and later
observed\cite{sick80} to exhibit scaling in the variable $y(q,\nu )$.  In the
simplest picture of $y$-scaling, the electron-nucleus cross section is divided
by the elastic nucleon cross section, leaving a universal function $F(y)$ which
is independent of $Q^2$ in the plane wave impulse approximation.  In the
scaling limit, $y$ can be interpreted as the nucleon's initial momentum along
the momentum transfer direction, and $F(y)$ is related to the nucleon's
momentum
distribution in the nucleus.  Thus, $y$ plays a similar role for nucleons in
a nucleus as $x$ does for quarks in a nucleon.

In the limit of high $Q^2$, the scaling variables $x$, $y$, and $\xi$ are
related.  In the parton model, $\xi$ replaces $x$ as the scaling variable when
the target mass is not neglected. At large $Q^2$, $\xi$ can also be expressed
as a function only of $y$ (with the leading scale-breaking term
$M^2/Q^2$)\cite{bwf92}.  There may also be a relationship between quasielastic
and inelastic scattering at more modest momentum transfers.
In the case of the free nucleon, Bloom and Gilman\cite{bloom71}
discovered that the resonance peaks in the structure function have the same
$Q^2$ behavior as the deep inelastic contribution when viewed as a function of
$\omega$', a modified version of the Bjorken scaling variable.  It was later
shown\cite{derujula} that this connection between the high $Q^2$ structure
function and the resonance form factors, called local duality, was expected
from perturbative QCD and should be valid for the nucleon elastic peak as well
as the resonance peaks if the structure function is analyzed in terms of $\xi$.
When the structure function is viewed as a function of $\xi$, the elastic
and resonance peaks have the same $Q^2$ behavior as the deep inelastic
structure function.  The elastic and resonance peaks decrease rapidly
with $Q^2$, but move to higher $\xi$, keeping a nearly constant strength
with respect to the deep inelastic structure function, which falls with
$\xi$.  Thus the strong $Q^2$ dependence of the higher twist effects (the
elastic
and resonance peaks) is removed when the structure function is averaged over a
range in $\xi$.  In the case of electron scattering from a nucleus, the Fermi
motion can perform this `averaging' of the structure function.  Thus, when
examining $\nu W_2^A$ as a function of $\xi$, scaling may be observed at
lower momentum transfers where $x$-scaling is not yet valid due to the
quasielastic contribution.

Scaling in inclusive scattering from nuclei (He, C, Fe and Au) was examined in
a
previous measurement\cite{day87,bwf92} for $Q^2$ values of 0.3-3.1 (GeV/c)$^2$.
 For these values of momentum transfer, where the quasielastic contribution
dominates the cross section, the data exhibit $y$-scaling for $y<0$ $(x>1)$.
The positive $y$ values represent the high energy-transfer side of the
quasielastic peak where the $y$-scaling breaks down due to the increasing
inelastic contribution at higher $Q^2$.  This same experiment examined $x$- and
$\xi$-scaling in the nucleus\cite{bwf92}.  For low values of $x$, the structure
function $\nu W_2$ is a function only of $x$, as predicted.  For values of $x$
near or above 1, scaling was not observed due to the contribution of
quasielastic scattering.  If one examines the structure function vs the
Nachtmann variable $\xi$, a scaling behavior is suggested.  At lower $\xi$, the
data are nearly independent of $Q^2$, while at higher $\xi$, the data approach
a universal curve.  More recent data at higher $Q^2$ show the same approach
to scaling for inclusive scattering from aluminum\cite{peb92}.  The beginning
of scaling in this region suggests that $\xi$-scaling is not just applicable to
deep inelastic scattering, but is also connected to quasielastic scattering and
$y$-scaling.  Here we examine scaling in the transition region from
quasielastic to deep inelastic scattering, to further study the connection
between $\xi$- scaling and $y$-scaling.

The data presented here are from the NE18
experiment\cite{tomthesis,carbonpaper}, a coincidence A(e,e'p) measurement
performed in End Station A at the SLAC Nuclear Physics Facility (NPAS).
Electron singles were recorded as well as the electron-proton coincidences and
this data was analyzed to measure the inclusive cross section.
Scattering was measured
from cryogenic liquid $^1$H and $^2$H targets and solid C, Fe, and Au targets
with beam energies of 2.02, 3.19, 4.21, and 5.12 GeV, at angles of 35.5$^o$,
47.4$^o$, 53.4$^o$, and 56.6$^o$ respectively ($Q^2$=1, 3, 5, and 6.8
(GeV/c)$^2$).  The scattered electrons were detected in the SLAC 1.6 GeV/c
spectrometer.  The pion rate in the spectrometer was up to 500 times the
electron rate for runs on Au at the highest $Q^2$.  A CO$_2$ gas \v{C}erenkov
counter and lead glass shower counter were used to eliminate the pions.  Tight
cuts were used in the final analysis, resulting in a pion rejection of 15000 to
1, while maintaining an electron efficiency of 90\%.

In order to extract the cross section, corrections for spectrometer acceptance,
detector efficiencies, data acquisition deadtimes, and radiative
corrections were applied to the data. The acceptance was
determined using a Monte-Carlo model of the spectrometer, and deadtime
corrections were measured on a run-by-run basis.  Radiative corrections were
applied using an iterative procedure following the formulae of Stein {\it et
al.}\cite{stein}, which are based on the work of Mo and Tsai\cite{motsai} and
Tsai\cite{tsai}.  A model cross section was radiated and compared to the data
to determine a smooth correction to the model cross section.  The `corrected'
model was then radiated again, and the procedure repeated until the radiated
model was consistent with the data.  The model dependence of the radiative
correction
procedure was tested by varying the initial model cross section.  We also
compared the radiatively corrected
cross sections calculated from runs using targets of
different thickness.  A final error of 3\% was assigned to the radiative
correction procedure.

Extracting the structure functions from the measured cross section without
performing a Rosenbluth separation requires a knowledge of the ratio of the
absorption cross sections for longitudinal and transverse virtual photons,
$R=\sigma _L/\sigma _T$.  However, the error in extracting $\nu W_2^A$ due to
uncertainty in $R$ is small for forward angles and for $R < 1$.  We have
assumed
$R = 0.5/Q^2$ with an uncertainty of 50\%, which is consistent with impulse
approximation predictions as well as a recent measurement\cite{peb92}.  This
leads to a worst case contribution to the uncertainty in $\nu W_2^A$ of $\pm$
3\%.  The scaling function $F(y)$ was extracted from the measured cross section
using the same method as Day {\it et al}\cite{day87} and
Potterveld\cite{dhpthesis}.

The extracted scaling function $F(y)$ for iron is shown in figure 1, along with
the previous SLAC NE3 data\cite{day87}.  While the $y < 0$ data exhibited
$y$-scaling for the previous data, the scaling clearly breaks down at high
$Q^2$ for all $y$ values measured ($y>-80$ MeV/c). The breakdown of
$y$-scaling is due to the transition from quasielastic scattering to inelastic
scattering.  To test $y$-scaling in this region, one must calculate and
subtract
off inelastic contributions to the cross section.  This introduces a model
dependence and can only be done reliably when the inelastic contributions do
not dominate the cross section. It is clear that in the case of inclusive
scattering, the applicability of $y$-scaling is limited to lower momentum
transfers, where quasielastic scattering dominates the cross section.

Figure 2 shows the measured structure function for iron and carbon as
a function of $x$.  Clearly the data do not scale in this range but the $Q^2$
dependence is decreasing as $Q^2$ increases.  The structure
function is nearly identical for all of the targets except deuterium, where
the smaller Fermi momentum causes a peak in the structure function near
$x=1$.  The larger Fermi momentum in the heavier nuclei washes out the
quasielastic peak, leading to a lower structure function near $x=1$.  The
difference between carbon and iron decreases as $x$ gets further from 1 and at
higher $Q^2$.
Table 1 gives the ratio of the structure functions for different targets at
each $Q^2$, for $0.95 < x < 1.05$, and the ratio of iron to deuterium for
$x = 1$.  The $Q^2$ behavior of the ratio to deuterium is consistent with the
behavior found for aluminum\cite{f&s92}. In figure 3, $\nu W_2$ is
plotted vs $\xi$ and an approach to scaling is observed. The new data are all
centered at $x=1$, but move to higher $\xi$ as $Q^2$ increases, lying on the
universal curve.

%To more
%easily see the scaling behavior, the scaling functions are shown vs. momentum
%transfer for fixed values of $y$, $x$, and $\xi$ (figure 4).  Here we clearly
%see the breakdown of $y$-scaling when $Q^2 > 3$ GeV$^2$/c$^2$, for $y$ of
%$-$0.08 GeV/c. We also see that while the structure function does not scale
%%for
%fixed values of $x$, we do see the approach to scaling for fixed $\xi$, with
%%an
%increase in $\nu W_2$ on the low $Q^2$ side of the elastic peak, and then a
%decrease to the high-$Q^2$ value, where inelastic scattering dominates.

To better understand the transition, we calculated the contributions due to
the different scattering processes using the convolution model of Ji
and Filippone\cite{ji90} with a Woods-Saxon spectral function, dipole
electric form factor, and the magnetic form factor of
Gari and Kr\"{u}mplemann\cite{gari85}.
Figure 4 shows the approach to scaling for fixed $\xi$ along with calculations
showing the quasielastic and deep inelastic contributions to the structure
function.  Also included is the NE11 data for aluminum \cite{peb92}, which are
in good agreement when the structure function is scaled by the number of
nucleons.  As a function of $Q^2$, we see an increase in the structure function
on the low $Q^2$ side of the quasielastic peak, and then a decrease to the
high-$Q^2$ value, where inelastic scattering dominates.  While the structure
function is not independent of $Q^2$ for a fixed $\xi$, it shows less $Q^2$
dependence than when viewed at a constant $x$.  More importantly, the measured
structure function has relatively little $Q^2$ dependence in the region of
transition from quasielastic to inelastic scattering, even though the
quasielastic contribution is falling rapidly with $Q^2$.  This is true for all
$\xi$ values measured, indicating a connection between the quasielastic and
inelastic cross sections, reminiscent of local duality in the nucleon.

To summarize, we have extracted the scaling function $F(y)$ and the structure
function $\nu W_2$ near $x=1$ for nuclei with $A$ ranging from 2$-$197 at $Q^2$
values from 1$-$7 (GeV/c)$^2$.  At the higher $Q^2$ values, $y$-scaling breaks
down for all measured values of $y$ as deep inelastic
scattering begins to dominate.  When examining $\nu W_2$, we do not
yet see scaling in $x$, but we do begin to see scaling
in the Nachtmann scaling variable $\xi$. This
suggests a connection between quasielastic and inelastic scattering,
similar to the case of local duality in the nucleon.  $\xi$-scaling may prove
to be a useful tool in understanding nuclear structure functions, but better
coverage in $\xi$ at the present $Q^2$ values, as well as higher $Q^2$
measurements (e.g. at CEBAF \cite{thesistobe}) are needed to fully understand
the scaling of the structure function, and the relation between $\xi$-scaling
and $y$-scaling.

This work was supported in part by the National Science Foundation under Grants
No. PHY-9014406 and PHY-9114958 (American), PHY-9115574 (Caltech), PHY-9101404
(CSLA), PHY-9208119 (RPI), PHY-9019983 (Wisconsin), and by the Department of
Energy under Contracts No. W-31-109-ENG-38 (Argonne), DE-FG02-86ER40269
(Colorado), W-2705-Eng-48 (LLNL), DE-AC02-76ER03069 (MIT), DE-AC03-76SF00515
(SLAC), DE-FG03-88ER40439 (Stanford).  RGM acknowledges the support of a
Presidential Young Investigator Award from NSF.  BWF acknowledges the support
of a Sloan Foundation Fellowship.

\begin{figure}
\caption{
$F(y)$ vs $y$ for iron for the present experiment and the previous NE3
measurement.  Errors in the new data (solid points) are dominated by a 4\%
systematic error, but are smaller than the points shown. }
\end{figure}

\begin{figure}
\caption{
$\nu W_2/A$ vs $x$ for iron (solid points) and carbon (hollow points). }
\end{figure}

\begin{figure}
\caption{
$\nu W_2^{Fe}$ vs $\xi$ for the present experiment and the NE3 measurement. }
\end{figure}

\begin{figure}
\caption{
$\nu W_2^{Fe}$ is plotted vs $Q^2$ at $\xi=0.85$. The lines are calculations of
the total (solid), quasielastic (dashed), and deep inelastic (dotted)
contributions to the structure function. }
\end{figure}

% \begin{table}
% \caption{}
% \label{}
% \end{tabular}
% \end{table}

\begin{table}
\squeezetable
\caption{Ratio of Structure Functions for different targets near $x = 1$}
\label{tab1}
\begin{tabular}{cccc}
Q$^2$ (GeV/c)$^2$ & C/Fe & Fe/Au & Fe/D \cr
\tableline
1.0&1.14$\pm$.05&1.11$\pm$.05&\cr
3.0&1.14$\pm$.05&1.19$\pm$.05&0.48$\pm$0.03 \cr
5.0&1.07$\pm$.05&1.16$\pm$.05&0.69$\pm$0.04 \cr
6.8&1.05$\pm$.05&0.96$\pm$.05&0.95$\pm$0.06 \cr
\end{tabular}
\end{table}

\narrowtext

\begin{references}
\bibitem[*] {atmaryland} Present address: University of Maryland,
College Park, Maryland 20742
\bibitem[\dag] {atboulder} Present address: University of Colorado,
Boulder, Colorado 80309
\bibitem[\ddag] {atmich} Present address: University of Michigan, Ann
Arbor, Michigan 48109
\bibitem[\S] {atcebaf} Present address: CEBAF, Newport News,
Virginia 23606
\bibitem[\parallel] {atillinois} Present address: University of Illinois
at Urbana-Champaign, Urbana, Illinois 61801
\bibitem[\P] {atanl} Present address: Argonne National Laboratory,
Argonne, Illinois 60439
\bibitem[**] {atvirgin} Present address: Virginia Union
University,
Richmond, Virginia 23220
\bibitem[\dag\dag] {atodu} Present address: Old Dominion University,
Norfolk, Virginia 23529
\bibitem[\ddag\ddag] {atpenn} Present address: University of
Pennsylvania, Philadelphia, Pennsylvania 19104
\bibitem[\S\S] {atcaltech} Present address: California Institute of
Technology, Pasadena, California 91125
\bibitem[\parallel\parallel] {atslac} Present address: SLAC, Stanford,
California 94309
\bibitem[\P\P] {atillinoistech} Present address: Illinois Institute
of Technology, Chicago, Illinois 60616
\bibitem[a] {atkent} Present address: Kent State University, Kent,
Ohio 44242
\bibitem[b] {atcolombia} Present address: Colombia University, New York, New
York 10027

\bibitem{hg76} H. Georgi and H. D. Politzer, Phys. Rev. D {\bf 14},
1829 (1976).
\bibitem{west75} G. B. West, Phys. Rep. {\bf 18}, 263 (1975).
\bibitem{sick80} I. Sick, D. B. Day, and J. S. McCarthy, Phys. Rev. Lett.
{\bf 45}, 871 (1980).
\bibitem{bwf92} B. W. Filippone, {\it et al.}, Phys. Rev. C {\bf 45},
1582 (1992).
\bibitem{bloom71} E. Bloom and F. Gilman, Phys. Rev. D {\bf 4}, 2901 (1971).
\bibitem{derujula} A. DeRujula, H. Georgi, and H. D. Politzer, Ann. Phys.
(N.Y.) {\bf 103}, 315 (1977). For a recent discussion of duality see
C. E. Carlson and N. C. Mukhopadhyay, Phys. Rev. D
{\bf 41}, 2343 (1990); Phys. Rev. D {\bf 47}, R1737 (1993).
\bibitem{day87} D. B. Day, {\it et al.}, Phys. Rev. Lett. {\bf 59},
427 (1987).
\bibitem{peb92} P. E. Bosted, {\it et al.}, Phys. Rev. C {\bf 46},
2505 (1992)
\bibitem{tomthesis} T. G. O'Neill, Ph.D. Thesis, California Institute of
Technology, 1994;  T. G. O'Neill {\it et al.}, paper submitted.
\bibitem{carbonpaper} N. C. R. Makins {\it et al.}, Phys. Rev. Lett. {\bf 72}
1986 (1994).
\bibitem{stein} Stein, {\it et al.}, Phys. Rev. D {\bf 12}, 1884 (1975).
\bibitem{motsai} L. W. Mo and Y. S. Tsai, Rev. Mod. Phys. {\bf 41}, 205 (1969).
\bibitem{tsai} Yung-Su Tsai, SLAC Report SLAC-PUB-848 (1971).
\bibitem{f&s92} L. L. Frankfurt, M. I. Strikman, D. B. Day, M Sargsyan, Phys.
Rev. C {\bf 48}, 2451 (1992).
\bibitem{gari85} M. F. Gari and W. Kr\"{u}mplemann, Z. Phys. {\bf A322},
   689 (1985).
\bibitem{ji90} Xiangdong Ji and B. W. Filippone, Phys. Rev. C {\bf 42},
R2279 (1990).
\bibitem{dhpthesis} D. H. Potterveld, Ph.D. Thesis, California Institute of
Technology, 1989.
\bibitem{thesistobe} CEBAF Proposal E89-008, D. B. Day and B. W. Filippone
Spokespersons.

%\bibitem{drell70} S. D. Drell and T. M. Yan, Phys. Rev. Lett. {\bf 24}, 181
%(1970).
%\bibitem{west70} G. B. West, Phys. Rev. Lett. {\bf 24}, 1206 (1970).
%\bibitem{llf87} L. L. Frankfurt and M. I. Strikman, Phys. Lett. B
%{\bf 183}, 254 (1987).
%\bibitem{jung88} H. Jung and G. A. Miller, Phys. Lett. B {\bf 200}, 351
%%(1988).
%\bibitem{jja83} J. J. Aubert {\it et al.}, Phys. Lett. B {\bf 123}, 275
%%(1983).
%\bibitem{bodek83} A. Bodek {\it et al.}, Phys. Rev. Lett. {\bf 51}, 534
%%(1983).
%\bibitem{arnold84} R. Arnold {\it et al.} Phys. Rev. Lett. {\bf 52}, 727
%%(1984).

\end{references}
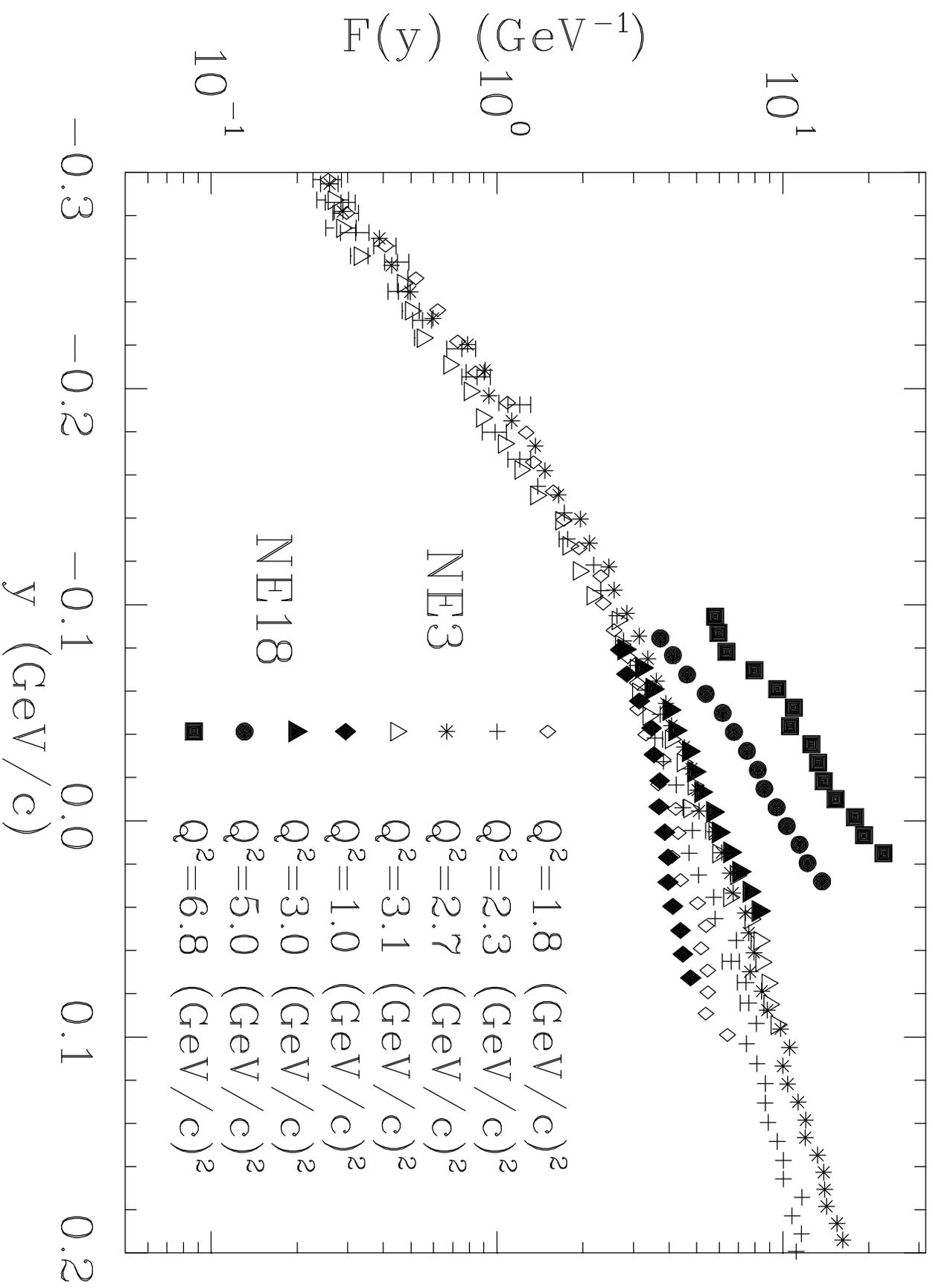
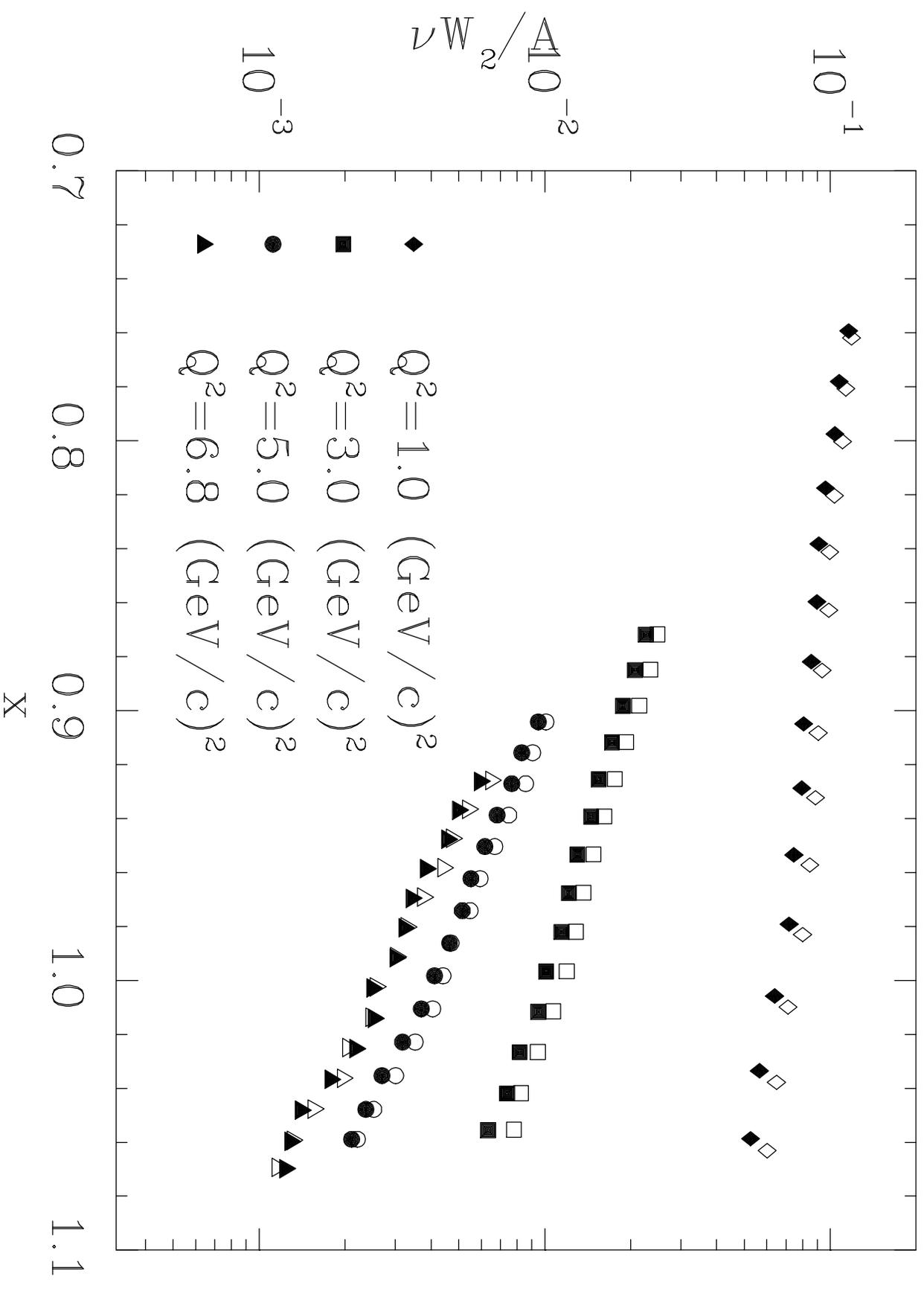
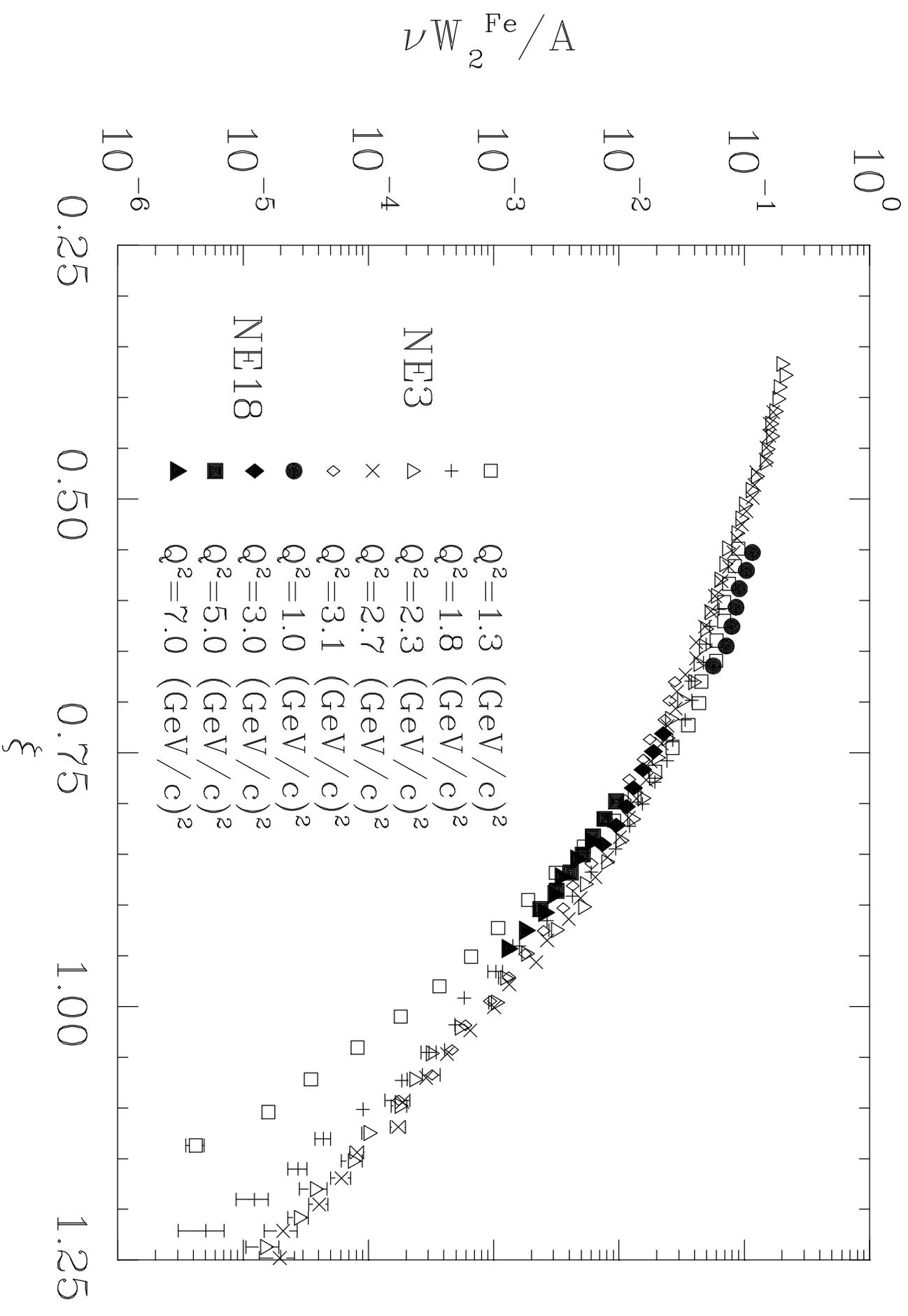
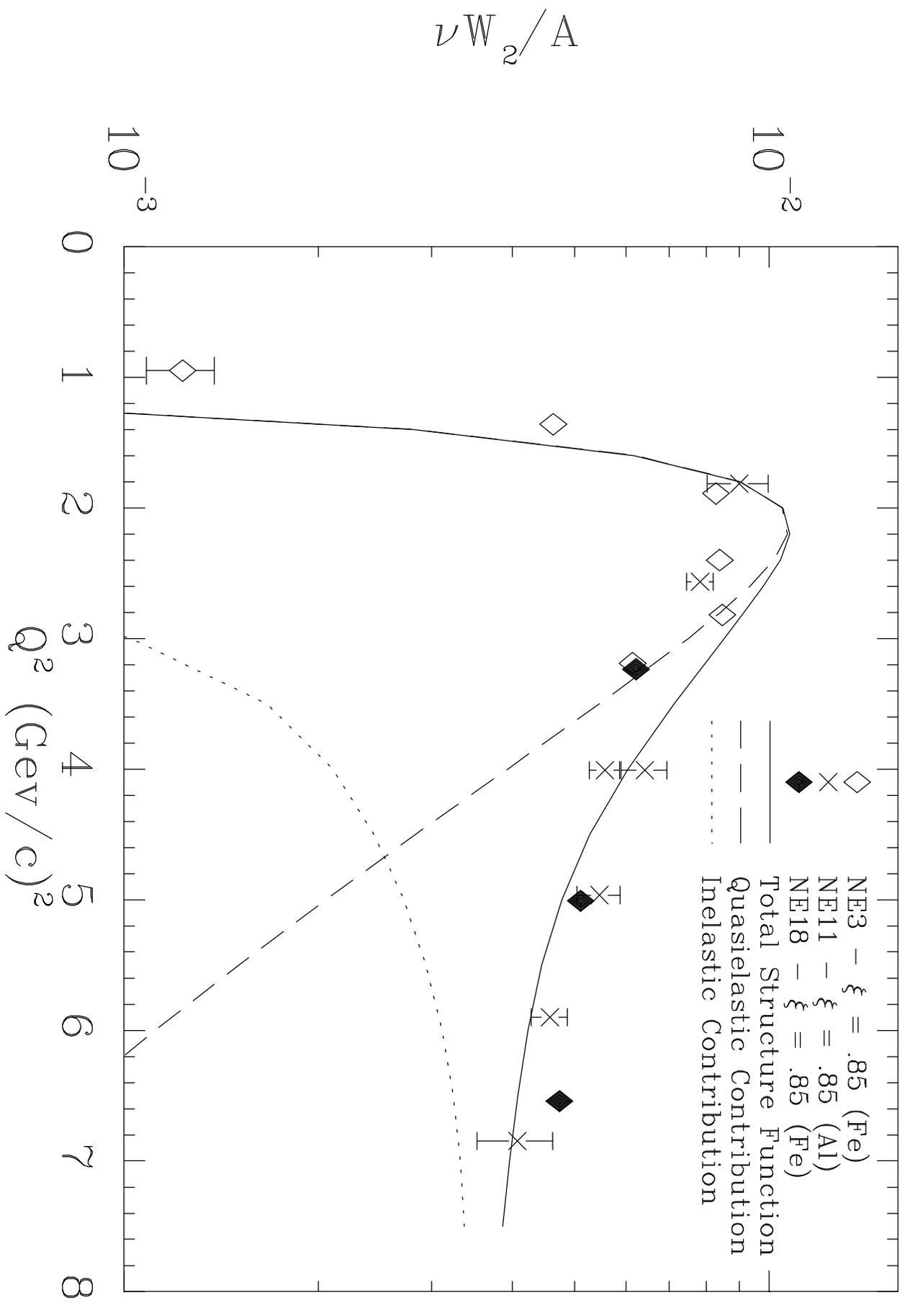
\end{document}